\documentclass{elsart}

\usepackage{graphicx} 
\usepackage{epsfig} 

\usepackage{subfigure}

\usepackage{amssymb}

\begin{document}

\begin{frontmatter}



\title{
Evidence for $\kappa$ Meson Production in $J/ \psi \to \bar K^*(892)^0 K^+ \pi^-$ Process 
}

\date{}
\begin{center}
\begin{small}

M.~Ablikim$^{1}$,           J.~Z.~Bai$^{1}$,            Y.~Ban$^{15}$,
J.~G.~Bian$^{1}$,           X.~Cai$^{1}$,               H.~F.~Chen$^{21}$,
H.~S.~Chen$^{1}$,           H.~X.~Chen$^{1}$,           J.~C.~Chen$^{1}$,
Jin~Chen$^{1}$
% ...
\\(BES Collaboration)\\
\vspace{0.2cm}
{\it
$^{1}$ Institute of High Energy Physics, Beijing 100049, People's Republic of China\\
% ...
$^{15}$ Peking University, Beijing 100871, People's Republic of China\\
% ...
$^{21}$ University of Science and Technology of China, Hefei 230026, People's Republic of China\\
% ...
}

\end{small}
\end{center}

\normalsize
%
%
\begin{abstract}
Based on 58 million  
BESII 
$J/ \psi$ events, the $\bar{K}^*(892)^0K^+\pi^-$ channel in 
$K^+K^-\pi^+\pi^-$ is studied. A clear low mass enhancement in the 
invariant  mass spectrum of $K^+\pi^-$ is observed. The low mass 
enhancement  does not come from background of other  $J/ \psi$ 
decay channels,  nor from phase space. Two independent partial 
wave analyses  have been performed. 
Both analyses favor  that the low mass enhancement is 
the $\kappa$, an isospinor scalar resonant state. 
The average mass and width of the $\kappa$ in the two analyses are 
878 $\pm$  23$^{+64}_{-55}$  MeV/$c^2$ and 
499 $\pm$ 52$^{+55}_{-87}$  MeV/$c^2$, respectively, 
corresponding to a pole at 
$ ( 841 \pm 30^{+81}_{-73} ) - i( 309 \pm 45^{+48}_{-72} ) $ MeV/$c^2$.
\end{abstract}

\end{frontmatter}


In the field of hadron spectroscopy, whether 
the low mass iso-scalar scalar meson, the $\sigma$, 
exists or not had been an important but controversial 
problem for many years. 
Recently, its evidence has been reported \cite{sigma2000}-\cite{sigma3} 
not only in $\pi\pi$  scattering,  
but also in various production processes. 
The $\sigma$  meson with a  mass around 600 MeV/$c^2$ 
and a broad width around  500 MeV/$c^2$ is, now, 
widely accepted \cite{PDG2004}.

The evidence for the $\sigma$ meson suggests the possibility of a
$\sigma$ nonet, \{$\sigma(600)$, $\kappa(900)$, $f_0(980)$, $a_0(980)$ \},
either in an extended linear sigma model realizing chiral
symmetry \cite{delbourgo}\cite{mishida} or 
in a unitarized meson model \cite{beveren}.
The $\kappa$ has been observed in analyses 
on $K\pi$ scattering phase shifts \cite{LASS} 
by several groups using a unitarized meson method \cite{beveren}, 
an interfering amplitude method \cite{sishida1} 
considering a repulsive background suggested by chiral symmetry, 
and a nonperturbative method \cite{black}-\cite{jawin} 
with an effective chiral Lagrangian. 
The observed  mass and  width values  are scattered in the ranges 
from 700 to 900 MeV/$c^2$ and 550 to 650 MeV/$c^2$, 
respectively, depending on the model used. 
Recently, 
a rather wider width around 800 MeV/$c^2$ was reported \cite{bugg} 
for the $\kappa$ in the analysis of $K\pi$ scattering phase shifts 
with a T-matrix method 
including a prescription for zero suppression. 
Also recently, an analysis of LASS data 
on $\pi K$ scattering phase shifts using a unitarization
method combined with chiral symmetry has found the $\kappa$ with a 
slightly lighter pole mass \cite{HQZHENG}.
However, some authors have found no evidence for 
the $\kappa$ \cite{lohs}-\cite{anisovich}. 
A criticism \cite{pennington1}  has been presented  for the existence of 
the $\kappa$ based on unitarity and universality arguments,  similar to
the case  of the $\sigma$ \cite{sishida2}\cite{pennington2}.

Here, it is to be noted that in $\pi\pi$ and $K\pi$ scattering, 
effects due to $\sigma$ and $\kappa$ 
production are, as a result of chiral symmetry, 
largely cancelled by those due to non-resonant background scattering, while
the cancellation mechanism does not necessarily work in the production 
process \cite{rf:psa}. Therefore, it is more suitable for 
the investigation of $\sigma / \kappa$ mesons to use 
the $\pi\pi/K\pi$ production process, 
which is parameterized independently 
of the scattering process \cite{scat-prod}.

Evidence for the $\kappa$ has been
reported, recently, in the production process by the E791 experiment at
Fermilab in the analysis of $D^+ \rightarrow K^-\pi^+\pi^-$ \cite{E791}.
Preliminary $\kappa$ results have been reported \cite{wuning2}
in the analysis of the $K\pi$ system produced in 
$J/\psi \rightarrow  \bar K^*(892)^0 K^+ \pi^-$ with BESI data.
The FOCUS experiment has presented \cite{link} evidence for the existence
of a coherent $K\pi$ S-wave contribution to $D^+ \rightarrow
K^-\pi^+\mu^+\nu$. CLEO \cite{anderson} has seen no evidence for 
the $\kappa$ in $D^0 \rightarrow K^-\pi^+\pi^0$. 
Preliminary results on the $\kappa$ have also been reported in 
analyses of BESII data \cite{wuning-nichidai}-\cite{WGLi-Had03}.

Here we report analyses of $\bar{K}^*(892)^0K^+\pi^-$ in 
$J/\psi \to K^+K^-\pi^+\pi^-$ to study the $\kappa$.  
Partial wave analyses (PWA analyses) have been performed 
in  $J/\psi \rightarrow \bar{K}^*(892)^0K^+\pi^-$ 
using 58 million $J/\psi$  decays obtained  with BESII 
at the BEPC (Beijing Electron Positron Collider) storage ring. 
The BESII detector is described in detail elsewhere \cite{besdetector}.
\\

\begin{figure}[htbp]
\epsfxsize=13.5cm \epsffile{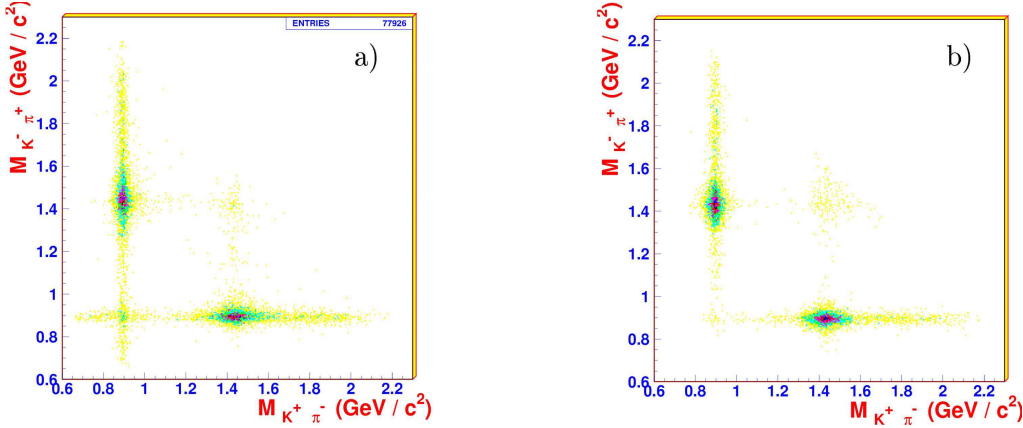}
\caption{
   {\footnotesize 
a) Scatter plot of $M_{K^+\pi^-}$ versus $M_{K^-\pi^+}$ 
for BESII data. 
b) Scatter plot of 
$M_{K^+\pi^-}$ versus $M_{K^-\pi^+}$ 
for Monte Carlo simulation of $J/\psi \rightarrow$ anything.
   }
}
\label{fig:1}
\end{figure}
In the event selection, four charged tracks with zero net
charge are required for each event. Every charged track should
have a good   helix fit in the Main Drift Chamber (MDC), 
and the polar angle $\theta$ of each track in the MDC 
must satisfy $|\cos \theta|<0.8$. 
The event must originate  near the collision point; 
tracks must satisfy 
$\sqrt{ x^2+y^2}   \,\le 2$ cm, $\mid z \mid \,\le 20$ cm, where $x$,
$y$ and $z$ are space coordinates 
of the point of closest approach of tracks to the beam axis.
Particle identification is performed using combined TOF 
and dE/dx information, 
and two identified kaons and two identified pions are required.

For a neutral track, it is required that  it should have hits 
in the Barrel Shower Counter (BSC), 
the number of layers hit should be greater than one, 
the shower starts before layer 6, 
and its deposited energy is more than 50 MeV. 
The angle between the photon emission direction and the shower 
development direction of the track in the BSC should be less than $30^\circ$.  
An event associated with a neutral track(s) is  rejected.

Surviving events are 
fitted kinematically (4C kinematic fits) under the hypotheses
$J/\psi \rightarrow K^+ K^- \pi^+ \pi^-$, $\pi^+ \pi^- \pi^+ \pi^-$,
$K^+ K^- K^+ K^-$, $K^+ \pi^- \pi^+ \pi^-$, and
$\pi^+ K^- \pi^+ \pi^-$.
It is required that $\chi^2_{4C}(K^+ K^- \pi^+ \pi^-)$ be less than 40. 
The sum of $\chi^2$ of the 4C kinematic fit, 
TOF, and dE/dx for $K^+ K^- \pi^+ \pi^-$ is required 
to be less than those for the 
$\pi^+ \pi^- \pi^+ \pi^-$,
$K^+ K^- K^+ K^-$, $K^+ \pi^- \pi^+ \pi^-$, and
$\pi^+ K^- \pi^+ \pi^-$ hypotheses.

In order to remove background from $J/\psi \to \phi \pi^+ 
\pi^-$, events with $|M_{K^+ K^-}-1.02| < 0.02$ GeV/$c^2$ are vetoed, and to 
remove $J/\psi \to \bar{K}^*(892)^0 K_S^0$ background, events  with 
$|M_{\pi^+ \pi^-}-0.497| < 0.04$ GeV/$c^2$ and $R_{xy} > 0.8$ cm are vetoed, 
where 
$M_{K^+ K^-}$ and $M_{\pi^+ \pi^-}$ are the invariant masses of $K^+ K^-$ 
and $\pi^+ \pi^-$, and $R_{xy}$ is the decay 
length of 
$K_S$ transverse to 
the beam axis.

Fig. \ref{fig:1}a shows the scatter plot of $M_{K^+ \pi^-}$ versus $M_{K^- \pi^+}$
after all above requirements.  Two bands of $\bar{K}^*(892)^0$ and
${K}^*(892)^0$ corresponding to $J/\psi \to \bar{K}^*(892)^0 K^+
\pi^-$ and $J/\psi \to {K}^*(892)^0 K^- \pi^+$, respectively, are
clearly seen. 
There is an accumulation of events 
in 
 the region where 
the two bands cross, which is not seen in the scatter plot of 
Monte Carlo $J/\psi \to anything$ events (shown in Fig. \ref{fig:1}b) obtained 
with the Lund-charm generator \cite{chenjc}, which is described below. 
This accumulation comes from $J/\psi$ decaying to 
$\bar{K}^*(892)^0$ (or ${K}^*(892)^0$) against a low mass enhancement.
%
%
\begin{figure}[thbp]
\epsfxsize=13.5cm \epsffile{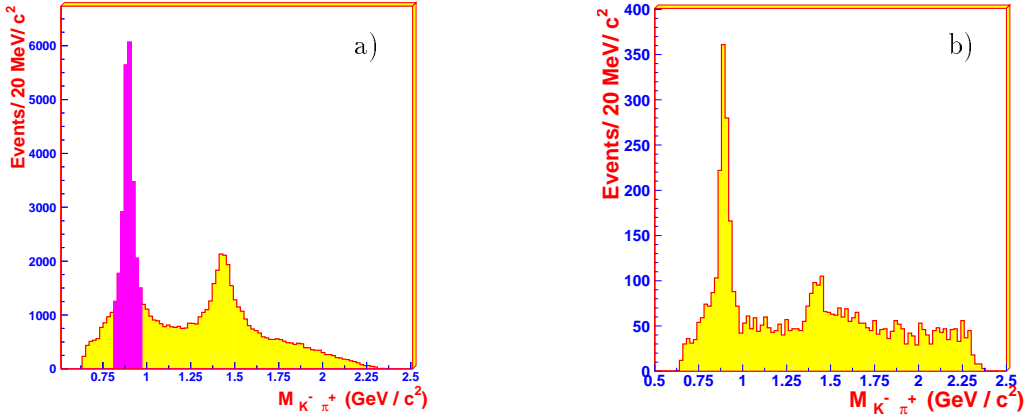}
\caption{
      {\footnotesize 
 a) Invariant mass spectrum for $K^-\pi^+$ after the final selection 
except the $\bar{K}^*(892)^0$ requirement. 
Events in the dark shaded area correspond 
to the data set in the present analyses. 
b) Recoil $K^-\pi^+$ 
 mass distribution against 
the low mass enhancement 
 ($M_{K^+\pi^-} < 0.8$ GeV/$c^2$). 
      }
}
\label{fig:2}
\end{figure}
%
The invariant mass spectrum of $K^- \pi^+$ is shown in Fig. \ref{fig:2}a, where
the $\bar{K}^*(892)^0$ peak is clearly seen.
The requirement 
0.812 GeV/$c^2$ $ < M_{K^- \pi^+}< $ 0.972 GeV/$c^2$ 
is imposed on
the $J/\psi \to K^+ K^- \pi^+ \pi^-$ sample to select $J/\psi \to
\bar{K}^*(892)^0 K^+ \pi^-$ events, the total number of which is 24674. 
The PWA analyses are applied to the selected 
$J/\psi \to \bar{K}^*(892)^0 K^+ \pi^-$ sample.

\begin{figure}[htbp]
\epsfxsize=14.cm \epsffile{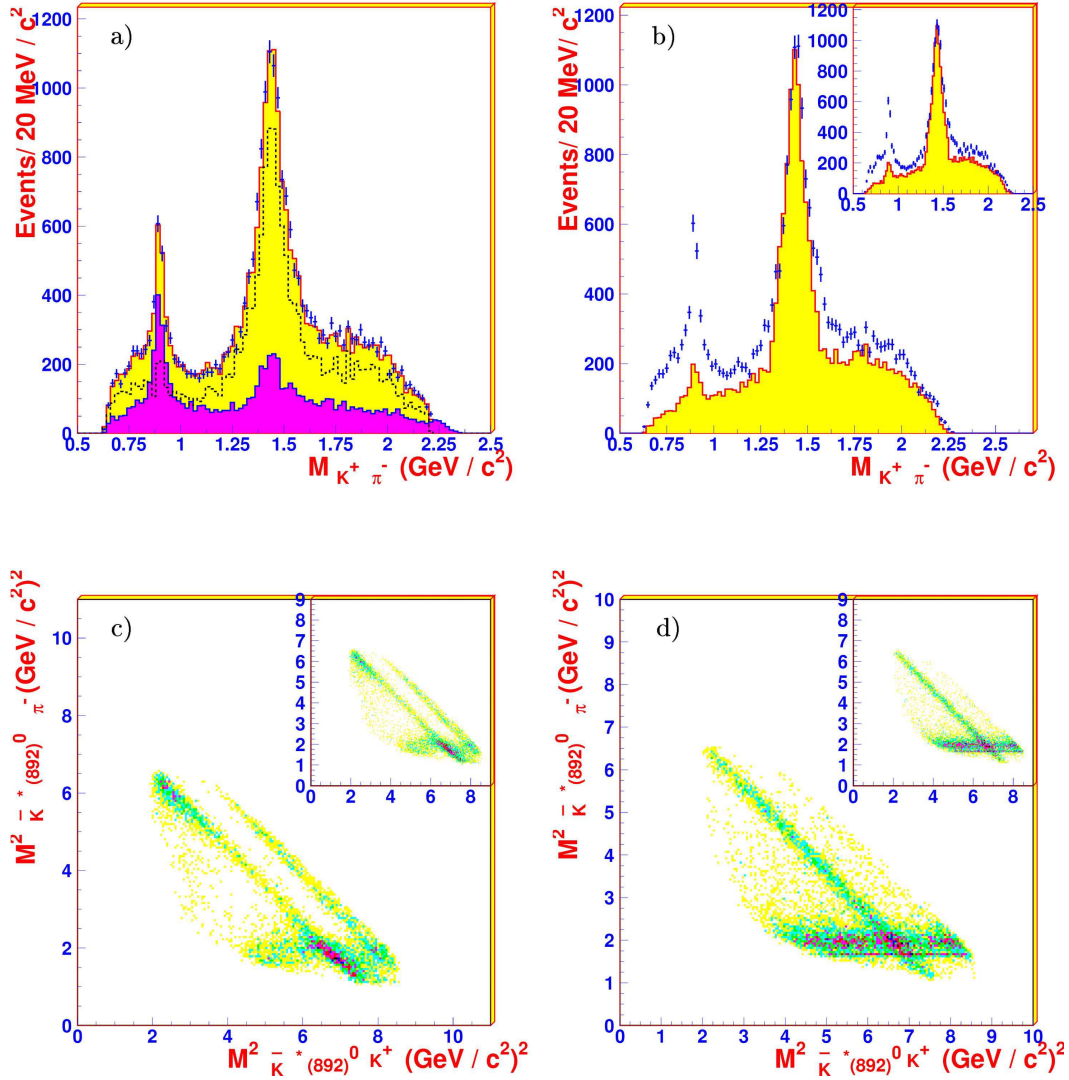}
\caption{
     {\footnotesize
a) Invariant mass spectrum of $K^+\pi^-$. 
The solid histogram is the data after the final selection, 
the dark shaded histogram is the $\bar{K}^*(892)^0$ side-band spectrum, 
and the dotted histogram is the invariant mass spectrum 
after side-band subtraction. 
Crosses with error bars show the charge conjugate state $K^-\pi^+$ from 
$J/\psi  \rightarrow K^*(892)^0K^-\pi^+$. 
b) Monte Carlo simulation compared with data. 
Crosses with error bars are data, and the solid histogram 
is Monte Carlo  simulation. 
c) Dalitz plot of data, and 
d) that of Monte Carlo simulation. 
Those of the charge conjugate state are shown as inserts in b), c), and d).
     }
}
\label{fig:3}
\end{figure}

After the $\bar{K}^*(892)^0$
mass requirement, the $K^+ \pi^-$ invariant mass distribution, shown as the
solid histogram in Fig. \ref{fig:3}a, has a clear 
${K}^*(892)^0$ peak, a peak around 1430 MeV/$c^2$, and a broad enhancement 
in the low mass region. The $K^- \pi^+$ mass distribution of 
the charge conjugate channel 
$J/\psi \to {K}^*(892)^0 K^- \pi^+$, denoted as crosses in Fig. \ref{fig:3}a, 
shows the same structures. 

Fig. \ref{fig:3}c is the Dalitz plot of 
$J/\psi \to \bar{K}^*(892)^0 K^+ \pi^-$, 
and the insertion is that of its charge conjugate channel.  In the
Dalitz plot, the two diagonal bands correspond to 
the low mass enhancement and the peak around 1430 MeV/$c^2$ 
in the $K^+ \pi^-$ invariant mass spectrum, and the horizontal 
band corresponds to $J/\psi \to K_1(1400) K$ and $K_1(1270) K$ with
$K_1(1400)$ and $K_1(1270)$ decaying to $\bar{K}^*(892)^0 \pi^-$.  
The clear top diagonal band in the Dalitz plot indicates 
that the low mass enhancement observed in this decay does not come from 
phase space since phase space events would be uniformly distributed 
in the Dalitz plot. 

Though this low mass enhancement overlaps with 
the narrow $ K^{\ast}(892)^0$,
it  can be clearly seen due to its broad structure.
The spectrum of $K^-\pi^+$ mass recoiling against the low mass enhancement 
($ M_{K^+\pi^-} < 0.8$ GeV/$c^2$) 
is shown in 
Fig. \ref{fig:2}b, where a clear $\bar{K}^*(892)^0$ peak can be seen. 
This means that the low mass enhancement is produced dominantly through 
$J/\psi$ decays associated with the $\bar{K}^{\ast}(892)^0$.
\\

$K^{\ast}(892)^0$ signals are clearly recognized in the $K^+ \pi^-$ spectrum 
against $\bar{K}^{\ast}(892)^0$ side-band events 
(the dark shaded histogram) in Fig. 3a. 
The side-bands events are taken from the $K^- \pi^+$ mass ranges 
directly neighboring to $\bar{K}^{\ast}(892)^0$ with 80 MeV/$c^2$ widths. 
After side-band subtraction, the $K^{\ast}(892)^0$ peak 
is suppressed appreciably in the invariant mass spectrum of
$K^+ \pi^-$,  as is shown in the same figure 
(the dotted histogram). 
This means that the $K^{\ast}(892)^0$ peak mostly 
comes from background processes. 
The main part of the broad low mass enhancement remains 
after side-band subtraction, which indicates that 
the broad low mass structure does not come from $J/\psi$ decay processes 
which do not contain $\bar{K}^*(892)^0$ in the final states.

The main background channels for $J/\psi \to \bar{K}^*(892)^0 K^+
\pi^-$ are studied through Monte Carlo simulation. More than 20
$J/\psi$ decay channels, including $J/\psi \to \gamma K^+ K^- \pi^+
\pi^-$, $\gamma \pi^+ \pi^- \pi^+ \pi^-$, $K^*(892)^{\pm} K^{\mp}$,
$K^*(892)^0 K_S$, and $\bar{K}^*(892)^0 K_S$ decays are generated
using uniform phase space generators, and no peak is produced in the
${K^+ \pi^-}$ mass spectrum. This means that the low mass broad
structure in the ${K^+ \pi^-}$ mass spectrum does not come from these
background channels. From the Monte Carlo simulation, we also see that
the background level in $J/\psi \to \bar{K}^*(892)^0 K^+ \pi^-$ is
about 1/7 of that in $J/\psi \to K^+ K^- \pi^+ \pi^-$.  Therefore, we
think $J/\psi \to \bar{K}^*(892)^0 K^+ \pi^-$ is a good place 
to study the $\kappa$. 
The background level ranges from 10 to 15 \% in this analysis. 

We also study 58 million Monte Carlo $J/\psi \to anything$ events 
which are generated using the Lund-charm model \cite{chenjc}.  
The generator is developed for simulating $J/\psi$ inclusive decay. 
In the models, charmonium decays into hadrons via quarks and gluons are 
simulated.
Quarks and gluons shower development and their 
hardronization are handled by the LUND string model. 
The model reproduces the main properties of hadronic 
events from $J/\psi$ inclusive decay. 
The process, $J/\psi \to K^*(892) \kappa$ is not included in the generator. 
Using the same 
selection criteria on this Monte Carlo sample as those for data, the scatter 
plot of $M_{K^+ \pi^-}$ versus $M_{K^- \pi^+}$ (Fig. \ref{fig:1}b), shows no 
accumulation of events 
around 
the region where the $K^*(892)^0$ and $\bar{K}^*(892)^0$ bands cross.
There is no corresponding broad low mass enhancement 
in the invariant mass spectrum of $K^+ \pi^-$ (or $K^- \pi^+$) 
recoiling against $\bar{K}^*(892)^0$ 
(or $K^*(892)^0$ for the charge conjugate channel) shown as the
shaded area in Fig. \ref{fig:3}b (or in the insertion of Fig. \ref{fig:3}b), 
and there is 
no diagonal band of the low mass enhancement in the Dalitz plot 
of $J/\psi \to \bar{K}^*(892)^0 K^+ \pi^-$ 
(or $J/\psi \to {K}^*(892)^0 K^- \pi^+$). 
Because the generator of this Monte Carlo simulation 
does not contain the process $J/\psi \to K^*(892) \kappa$, the 
Monte Carlo shows different structures at lower $K \pi$ mass region 
from those of data. This difference just means that the low mass 
enhancement is not due to backgrounds coming from other $J/\psi$ 
decay channels.

In the PWA of $J/\psi \to \bar{K}^*(892)^0 K^+
\pi^-$, the $\bar{K}^*(892)^0$ is treated as a stable particle. 
After integrating 
over the $K^- \pi^+$ angular information, no interference effect 
is expected between the charge conjugate processes, 
$J/\psi \to \bar{K}^*(892)^0 K^+ \pi^-$ and ${K}^*(892)^0 K^- \pi^+$, 
if the solid angle coverage of the detector is $4\pi$. 
We examined the interference effect taking into account 
the detector acceptance and the width of the $\bar{K}^*(892)^0$ 
using Monte Carlo simulation and find 
that the interference effect in the cross region of the two $K^*(892)^0$
bands in the scatter plot is negligibly small, 
and that the interference between 
$\bar{K}^*(892)^0 K^+ \pi^-$ and $\rho K \bar K$ (or $K^*_J(1430)K \pi$) 
is also 
negligible. 
We ignore these interferences in the PWA analyses. 
\\

\begin{figure}[htbp]
\epsfxsize=13.0cm \epsfysize=6.5cm \epsffile{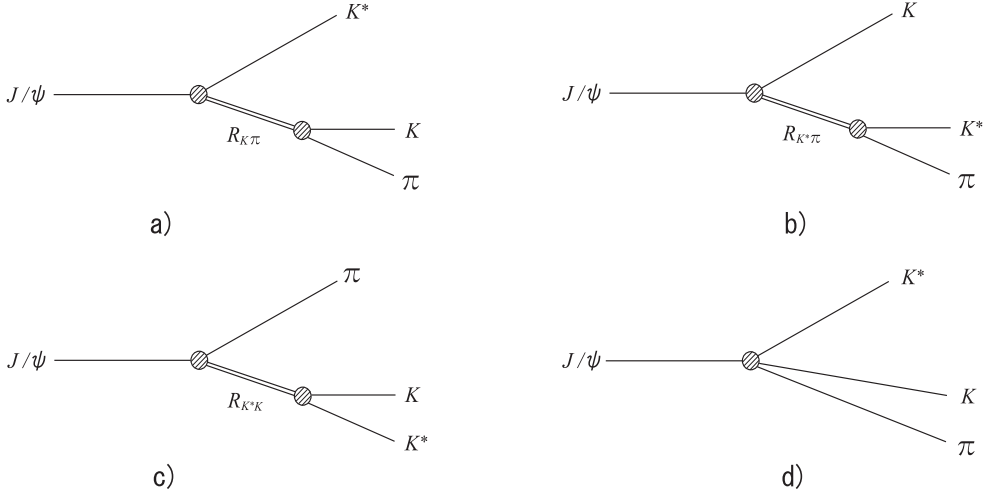}
\caption{
     {\footnotesize
Decay mechanisms for 
a) $J/\psi \rightarrow K^*(892)R_{K\pi}$, 
b) $J/\psi \rightarrow KR_{K^*\pi}$, 
c) $J/\psi \rightarrow \pi R_{K^*K}$, and 
d) $J/\psi \rightarrow K^*K\pi$. 
$R_{K \pi}$ , $R_{K^* \pi}$, and $R_{K^* K}$ are intermediate resonant states 
decaying into $K \pi$, $K^* \pi$ and $K^* K$, respectively. 
$K^*$ denotes $K^{*}(892)^0$. 
     } 
} 
\label{fig:4}
\end{figure}
We consider the $K^*(892)^0 K \pi$ channel in $J/\psi$ decays,
with well defined quantum states recoiling against the $K^*(892)^0$, 
is suitable to study the $\kappa$ in the $K\pi$ system. 
The $K^*(892)^0K\pi$ system simplifies the PWA, and
backgrounds can be treated easily. 
To simplify the analysis, one of the charge conjugate states, 
$\bar{K}^*(892)^0 K^+\pi^-$, is used, and it has enough events 
for this analysis. 
As mentioned above, 
the analysis of this channel is not affected by the charge conjugate state, 
$K^*(892)^0K^-\pi^+$, since the interference between them 
is found to be negligibly small.

Four decay mechanisms considered in
our analyses are shown in Fig. \ref{fig:4}. 
These are $J/\psi \rightarrow
K^*R_{K\pi}$ (Fig. \ref{fig:4}a), 
$J/\psi \rightarrow KR_{K^*\pi}$ (Fig. \ref{fig:4}b), 
$J/\psi\rightarrow \pi R_{K^*K}$ (Fig. \ref{fig:4}c) and 
$J/\psi \rightarrow K^*K\pi$ (Fig. \ref{fig:4}d), where
$R_{K\pi}$, $R_{K^*\pi}$ and $R_{K^*K}$ are intermediate resonant states 
decaying into $K\pi ~ (\kappa , K_0^*(1430), K_2^*(1430), K_2^*(1920))$, 
$K^*\pi ~ (K_1(1270), K_1(1400))$ and $K^*K ~ (b_1(1235))$, respectively. 
(Here, $K^*$ denotes $K^{*}(892)^0$.)
The independent PWA analyses by Method A and Method B have been performed on 
the $\bar{K}^{*}(892)^0K^+\pi^-$ channel. 
The same data set is used in both analyses.

Method A is based on the covariant helicity amplitude analysis \cite{wuning3}.
The maximum likelihood method is utilized in the fit.
The fit is performed by using Breit-Wigner 
parameterizations 
with an $s$-dependent width, $\rho(s)$, (See eq. (\ref{eq:ma1}) below.) 
for the low mass  enhancement and with constant widths for 
the other intermediate states. 
 Fits are also  performed  using  two other parameterizations of
 constant  width  
and  of a width for the $\kappa$ using a unitarization approach with chiral
 symmetry \cite{HQZHENG} as follows; 
\begin{eqnarray}
       \label{eq:ma1}
 BW &\propto& 1/(M^2-s-i \sqrt{s} \Gamma_{\kappa}(s))  ~ ,~ 
   \Gamma_{\kappa}(s)   =   g^2_{\kappa}\cdot k_K/8\pi s ~ , 
\\
       \label{eq:ma2}
 BW &\propto& 1/(M^2-s-i M \Gamma_{const})  ~ ,~ 
\end{eqnarray}
and 
\begin{eqnarray}
       \label{eq:ma3}
 BW &\propto& 1/(M^2-s-i \sqrt{s} \Gamma_{\kappa}(s))  ~ ,~ 
   \Gamma_{\kappa}(s)   =   \alpha_{\kappa} k_K     .
\end{eqnarray}
Eq. (\ref{eq:ma1}) is for Method A, 
eq. (\ref{eq:ma2}) is for Method A-1, and 
eq. (\ref{eq:ma3}) is for Method A-2.

 The broad low mass enhancement is fit by a $0^+$ resonance. 
The possibility that its spin-parity is $1^-$, $2^+$,
$\cdots$ is excluded by at least 
10$\sigma$. 
If the kappa is removed from the fit, the log-likelihood
becomes worse by 294. So, its
 statistical significance is above 20$\sigma$.
This iso-spinor scalar resonance is considered to be
the $\kappa$ particle. 
The above parameterizations are  tried for the $\kappa$. 
Though these parameterizations have different behavior, 
the quality of the fits given by them is almost the same. 
This is because there are many resonances with interferences between them, 
and changes in one can be compensated by changes in the others while 
keeping the total contribution unchanged.
Though mass and width parameters given by these 
parameterizations are somewhat different, the shapes of 
the $\kappa$ obtained by these parameterizations 
are similar, and the pole positions are  close to each other.

The biggest peak at about 1430 MeV/$c^2$ in the $K^+ \pi^-$ spectrum 
is relatively complex, containing $0^+$, $1^-$, and $2^+$ components. 
The $0^+$ and $2^+$ components are identified to be
$K_0^*(1430)$ and $K_2^*(1430)$, respectively, and 
their masses and widths determined from the 
fit are consistent with PDG values \cite{PDG2004}. 
A $1^-$ is included in the fit in
this region. Changes by removing it are considered
in the systematic uncertainties.
In the higher $K^+ \pi^-$ mass region, a broad resonance is needed.
It is found that different treatments of 
it have little influence on the mass and width of the $\kappa$.

$K_1(1270)$ and $K_1(1400)$ are used to fit the 
enhancement near threshold of the $\bar{K}^*(892)^0$$ \pi^-$ 
spectrum, and $b_1(1235)$ is used to improve the fit 
in the $\bar{K}^*(892)^0 K^+$ spectrum. 
Because the mass of $b_1(1235)$ is below $\bar K^*(892)^0 K$ threshold, 
only the tail of $b_1(1235)$ affects 
the fit of the $\bar K^*(892)^0 K$ spectrum. 
Changes caused by removing the $b_1(1235)$ are 
included in the systematic uncertainties.

In the PWA, the backgrounds from the charge conjugate channel 
$J/\psi \to {K}^*(892)^0 K^- \pi^+$ and from $J/\psi \to \bar K^*(892)^0 K_S$
and $K_S \to \pi^+ \pi^-$ where $\pi^+$ is misidentified with $K^+$ 
are fitted by non-interfering amplitudes, 
and the background from other $J/\psi$ decay channels, 
including $J/\psi \to \rho K \bar{K}$ are fitted by 
non-interfering phase space. 
%
\begin{table}[hbpt]
 \caption{
     {\footnotesize
 Masses, widths and pole positions of the $\kappa$ obtained by
Methods A (eq. (\ref{eq:ma1})), B (eq. (\ref{eq:mb})),  
A-1 (eq. (\ref{eq:ma2})), and A-2 (eq. (\ref{eq:ma3})). 
The mass and width values averaged for  A and B
are given.  The first term errors are statistical ones and the
second show systematic uncertainties estimated in the analyses.
     }
  }
 \begin{center}
\begin{tabular}{|ll|l|l|l|l|} \hline
             & Mass (MeV/$c^2$)    &  Width (MeV/$c^2$)  & Pole position (MeV/$c^2$)  \\ \hline
Method A & $874\pm25^{+12}_{-55}$ & $518\pm65^{+27}_{-87}$ & $(836\pm38^{+18}_{-87})-i(329\pm66^{+28}_{-46})$   \\ 
Method B & $896\pm54^{+64}_{-44}$ & $463\pm88^{+55}_{-31}$ &
 $(865\pm70 ^{+78}_{-60})-i(271\pm67 ^{+44}_{-23})$ \\
Average  & $ 878\pm  23  ^{+64}_{-55}$  
         & $ 499\pm  52^{+55}_{-87}$ 
         & $    ( 841\pm   30  ^{+81}_{-73} )  
             - i( 309\pm  45  ^{+48}_{-72} )$ \\ \hline 
Method A-1 & $745\pm26^{+14}_{-91}$& $622\pm77^{+61}_{-78}$ & $(799 \pm37^{+16}_{-90})-i(290\pm33^{+25}_{-38})$   \\ 
Method A-2 & $1140\pm39^{+47}_{-80}$ & $1370\pm156^{+406}_{-148}$&$(811\pm74^{+17}_{-83})-i(285\pm20^{+18}_{-42})$ \\\hline 
\end{tabular}
\end{center}
\end{table}
%
The uncertainty of the background level is considered, 
and its influence on the mass and width of the $\kappa$ is estimated 
and is put into the systematic errors.

Since there exist broad $K^+ \pi^-$ resonances around 1430 MeV 
and those broad resonances and other background processes 
$K^*(892), \ K_S$, and PS contribute in the $\kappa$ region 
, and because of the complicated 
interferences between $\kappa$ and other resonances, 
different solutions are obtained in the fits.
These solutions give almost the same fit quality and almost the same 
pole position of $\kappa$. 
 Uncertainties coming from the multi-solutions on the $\kappa$ parameters 
are included in the systematic errors.

\begin{figure}[hbpt]
\begin{center}
\epsfxsize=11.cm \epsfysize=17.cm \epsffile{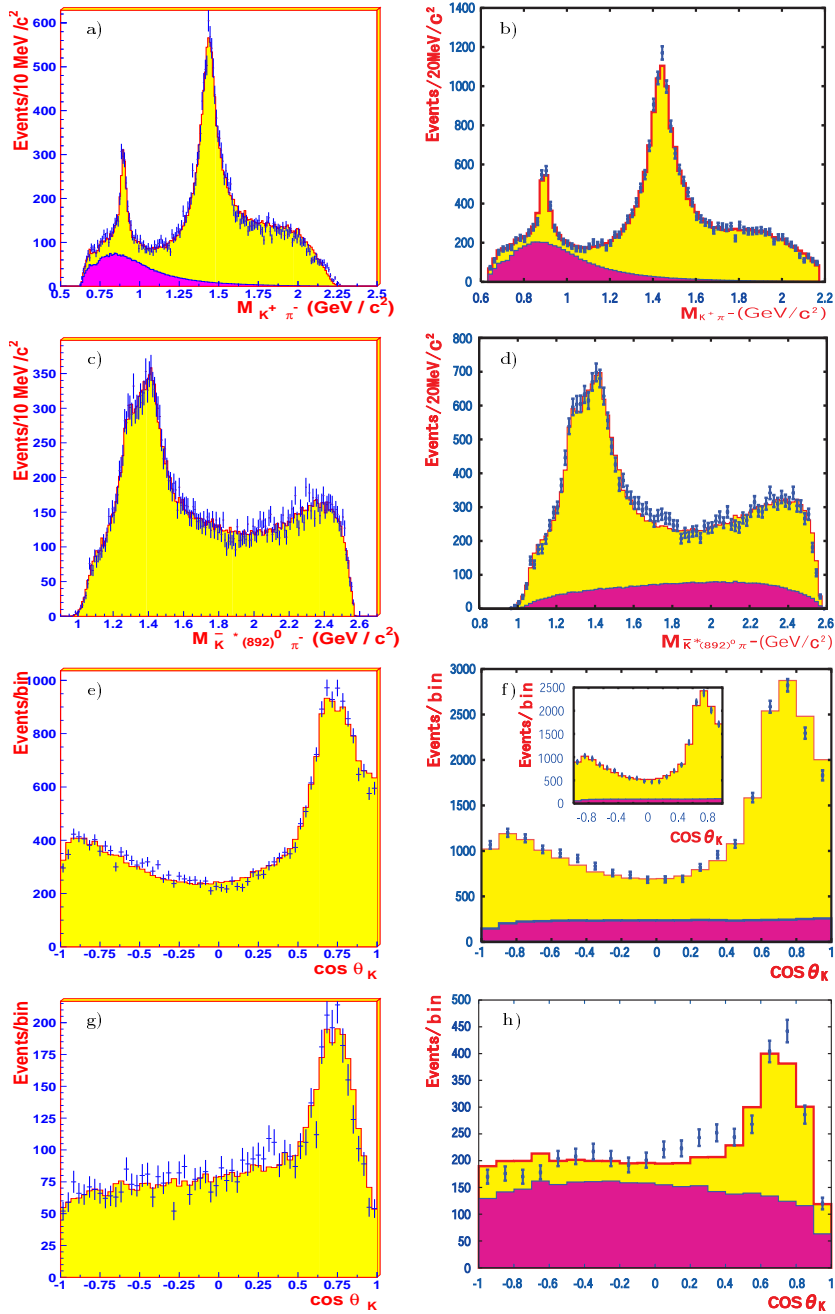}
\end{center}
\caption{
     {\footnotesize
Analysis results by Method A. 
a) The $K^+\pi^-$ invariant mass spectrum. 
c) The $\bar K^*(892)^0 \pi^-$ invariant mass spectrum. 
e) The $K^+$ angular distribution  in the $K^+\pi^-$ center of mass 
system  for the whole $K^+\pi^-$ mass region, and 
g) that for the $K^+\pi^-$  mass region below 1.0 GeV/$c^2$. 
Crosses with error bars are data. 
Solid histograms show fit results, and the dark shaded histogram in a) is 
the contribution from the $\kappa$.
Analysis results by Method B. 
b) The $K^+\pi^-$ invariant mass spectrum. 
d)The $\bar K^*(892)^0 \pi^-$invariant mass spectrum. 
f) The $K^+$ angular distribution in the $K^+\pi^-$ center of 
mass system  for the whole $K^+\pi^-$ mass region with that for 
above 1.0 GeV/$c^2$ in the insertion, and 
h) that for the $K^+\pi^-$   mass region below 1.0 GeV/$c^2$. 
Crosses with error bars are data. 
Solid histograms are fit, and 
dark shaded histograms are contributions from the $\kappa$.
     }
}
\label{fig:5}
\end{figure}

The fit obtained in the analysis is shown by the solid histogram 
in  Fig. \ref{fig:5}a for the $K^+\pi^-$   invariant mass  spectrum. 
The data are shown by crosses with error bars. 
The contribution of the $\kappa$ is shown by the dark shaded histogram. 
The fit of the $\bar K^*(892)^0 \pi^-$ invariant mass spectrum is shown 
by the solid histogram in  Fig. \ref{fig:5}c. 
The fit for the angular distribution of the whole $K^+\pi^-$ mass 
region is shown by the solid histogram in Fig. \ref{fig:5}e, and 
that for the  angular distribution of the $K^+ \pi^-$ mass below 1.0 GeV/$c^2$ 
is shown by the histogram in Fig. \ref{fig:5}g. 

The values for Breit-Wigner parameters of mass and width and 
those for the pole position for the $\kappa$ obtained 
in Method A are given in Table 1.
Systematic uncertainties in the mass and width 
of the $\kappa$ include the changes 
from 1$\sigma$ variations of the masses and widths of the other resonances, 
different treatments of background, and uncertainties from the fitting.

Mass and width parameters of intermediate resonances 
and of those background processes in the fit of Method A are tabulated in Table 2.
Errors shown in the table are only statistical.

\begin{table}[htp]
\label{tab:2}
\begin{center}
\caption { 
Mass and Width values of intermediate resonance states and those of background processes in the fits of Method A and Method B. BG and PS denotes background and phase space, respectively.
}
\begin{tabular}{|l|c|c|c|c|}
\hline
           & \multicolumn{2}{c|}{Method A} 
           & \multicolumn{2}{c|}{Method B} \\ \hline 
  Resonances  & mass (MeV/$c^2$) &  $\Gamma$ (MeV/$c^2$) 
           & mass (MeV/$c^2$) &  $\Gamma$ (MeV/$c^2$) \\ \hline 

$\kappa \rightarrow K \pi $      & 874$\pm$25  & 518$\pm$65
                                 & 896$\pm$54  & 463$\pm$88\\
\hline

$K^*_0(1430) \rightarrow K \pi $ & 1436 $\pm 46$  & 348  $\pm 55$ 
                                 & 1448 $\pm 37$  & 200$^{1)}$ \\
\hline

$K^*_2(1430) \rightarrow K \pi $ & 1434 $\pm  7$   &  101  $\pm 15$ 
                                 & 1443 $\pm 48$   &  97   $\pm 16$ \\
\hline

X(1430) ($1^-$) $ \rightarrow K \pi$& 1426 $\pm 6$ & 152 $\pm 16$ 
                                 & \multicolumn{2}{c|}{not included$^{2)}$} \\ 
\hline

$K^*_2(1920) \rightarrow K \pi $ & 1914 $\pm 36$ & 590 $\pm 110$ 
                                 & 1911 $\pm 24$ & 244 $\pm  95$ \\
\hline

$K_1(1270) \rightarrow K^*(892) \pi $ & 1279 $\pm 10$   & 131  $\pm 21$ 
                                 & 1271 $\pm 12$   & 104  $\pm 33$ \\
\hline

$K_1(1400) \rightarrow K^*(892) \pi $ & 1418 $\pm  8$   & 152 $\pm 16$ 
                                 & 1429 $\pm 31$   & 200$^{3)}$ \\
\hline

$b_1(1235) \rightarrow K^*(892) K $   &  fixed$^{4)}$  &  fixed$^{4)}$
                                 &   fixed$^{4)}$ &  fixed$^{4)}$  \\
\hline

Direct $\bar{K}^*(892)^0 K \pi$  & \multicolumn{2}{c|}{not included$^{2)}$} 
                                 & \multicolumn{2}{c|}{included$^{5)}$} \\ 

\hline
\hline

$ K^*(892)$ BG $ \rightarrow K \pi $    & fixed$^{4)}$  & fixed$^{4)}$
                                        & fixed$^{4)}$  & fixed$^{4)}$ \\
\hline

$K^*(1410)$ BG $ \rightarrow K \pi $   & \multicolumn{2}{c|}{not included$^{2)}$} 
                                       & fixed$^{4)}$  & fixed$^{4)}$ \\
\hline

$K_S$ BG                         &  715$^{6)}$   &  46$^{6)}$
                                 &  715$^{6)}$   &  46$^{6)}$ \\
\hline

$ K^*(892) K \pi$ PS BG          & \multicolumn{2}{c|}{included$^{5)}$} 
                                 & \multicolumn{2}{c|}{included$^{5)}$}  \\
\hline

\end {tabular}
\end{center}
1) Bound for the lower limit which is set to be consistent with that of the PDG tables. \\
2) The process is not included in the fit. \\
3) Bound for the upper limit which is set to be consistent with that of the PDG tables. \\
4) The value is fixed to that of the PDG tables. \\
5) The process is included in the fit. \\
6) The value is parameterized by the Monte Carlo simulation and fixed in the fit.
\end{table}

Method B uses the VMW (Variant Mass and Width) method, 
a covariant field theoretical approach consistent with 
generalized unitarity \cite{scat-prod}. 
In this method,  the total amplitude is expressed as a coherent  sum of
respective amplitudes, corresponding to the relevant processes  of strong
interactions  among all color-singlet hadrons.  
As the bases of the $S$-matrix for the strong interaction, 
a residual interaction of QCD, all unstable (or resonant) 
as well as stable hadrons which are to be color-singlet bound states 
of quarks, anti-quarks and gluons are to be included. 
The propagator of a resonant particle is given by 
the conventional Feynman propagator with substitution of 
i$\epsilon$ by $i\sqrt{s}{\Gamma(s)}$.

For the scalar $K\pi$-resonant particles, $R_{K \pi}$'s are 
the $\kappa$ and $K_0^*(1430)$. 
The Lagrangian of strong interaction, ${\mathcal L}_{\rm S}$, describing 
the process 
in Fig. \ref{fig:4}a is taken to be the most simple form.
This form and the corresponding decay amplitude ${\mathcal F}_S$ are given by
\begin{eqnarray}
 {\mathcal L}_S &  = & \sum_{R=\kappa ,K_0^*} ( \xi_R \psi_\mu K^*_\mu R + g_R RK\pi ), \nonumber \\
{\mathcal F}_S & = &  S_{h_\psi h_{K^*}} \sum_{R=\kappa ,K_0^*} 
r_R e^{i\theta_R} \Delta_R (s_{K\pi}), \nonumber \\
\Delta_R(s_{K\pi}) &=&
\frac{m_R\Gamma_R}{m_R^2-s_{K\pi}-i\sqrt s_{K\pi}\Gamma_R(s_{K\pi})},
\label{eq:mb}
\end{eqnarray}
where $\Delta_R(s_{K\pi})$ is the Breit-Wigner formula with 
$\Gamma_R(s_{K\pi})= pg_R^2/(8\pi s_{K\pi})$, describing the decay of
$R=\kappa$ and $K_0^*(1430)$, and $S_{h_\psi h_{K^*}}$ is the helicity amplitude
$S_{h_\psi h_{K^*}}\equiv \epsilon_\mu^{(h_\psi )} 
\tilde\epsilon_\mu^{(h_{K^*})}$.
$S_{h_\psi h_{K^*}} r_R e^{i\theta_R}$ describes the $S$-matrix
element $_{out}\langle RK^* | J/\psi \rangle_{in}$, where $e^{i\theta_R}$
parametrizes the rescattering phase of $_{out}\langle RK^* |$. This form of 
${\mathcal F}_S$ is consistent with the generalized unitarity of the $S$-matrix.

The decay amplitudes through the tensor $R_{K\pi}$,  $R_{K^*\pi}$, 
and $R_{K^*K}$, denoted as 
${\mathcal F}_D$, ${\mathcal F}_{K_1}$, and ${\mathcal F}_{b_1}$, respectively, 
are obtained in a similar manner. 
The direct $K\pi$ production amplitude is taken to be
${\mathcal F}_{\rm dir}= S_{h_\psi h_{K^*}} r_{K\pi} e^{i\theta_{K\pi}}$,
which is from ${\mathcal L}_{\rm dir}\sim \xi_{K\pi} \psi_\mu
K^*_\mu K\pi$. The total amplitude ${\mathcal F}$ is given by the sum of all these amplitudes,
\begin{eqnarray}
{\mathcal F} &=& {\mathcal F}_S+{\mathcal F}_D+{\mathcal F}_{K_1}+{\mathcal F}_{b_1}
+{\mathcal F}_{\rm dir}.
\end{eqnarray}
We also consider the background processes coming from $J^P=1^-$ $K^{*}(892)$ 
and $K^{*}(1410)$ (decaying into $K^+ \pi^-$), from $K_S$, and from 
phase space  $K^*(892) K\pi$, 
which are described by the amplitudes
incoherent with the above ${\mathcal F}$. The details are 
described elsewhere \cite{komada-nichidai}.
The treatments of resonances and background processes in this method 
are the same as those in Method A, 
except for the $K^*(1410)\bar{K}^*(892)$ and $K^*(892) K\pi$ processes, 
as explained below.

The least $\chi^2$  method is used 
for the fitting of the mass distribution  of $K^+\pi^-$, that of
$\bar{K}^*(892)^0 \pi^-$, and the $K^+$ angular distribution 
in the $K^+\pi^-$ system. 
The results obtained in this analysis  are shown by the solid histograms for 
the $K^+\pi^-$  invariant mass  spectrum in Fig. \ref{fig:5}b and  for the
$\bar{K}^*(892)^0 \pi^-$ mass spectrum in Fig. \ref{fig:5}d.
The data for the analysis are shown by  crosses  with error bars. 
The contributions of the $\kappa$ are shown by 
the dark shaded histograms 
superimposed on Figs. \ref{fig:5}b and \ref{fig:5}d.
The results  for the $K^+$ angular distributions in the whole $K^+\pi^-$ 
mass  region  and below 1.0 GeV/$c^2$ are shown by the solid histograms in 
Figs. \ref{fig:5}f and \ref{fig:5}h, respectively, and 
that for the mass region above 1.0 GeV/$c^2$ is inserted in 
Fig. \ref{fig:5}f. 
The contributions of the $\kappa$ are also shown by the dark shaded histograms 
in the figures.

The values for Breit-Wigner parameters of mass and width and 
pole position for the $\kappa$ obtained in the analysis of 
Method B are given in Table 1. 
Uncertainties are estimated on the same items as in Method A.  Method B
takes  the direct $\bar{K}^*(892)^0K^+\pi^-$ process to be coherent 
and phase space background contribution to be incoherent. 
The contribution of the latter is estimated using 
the results obtained in Method A.  
The uncertainties of it are included in the systematic errors of the 
$\kappa$ parameters.
The $\bar{K}^*(892)^0K^*(1410)$ amplitude is taken as an incoherent background 
process. 
The $K^*(892)^0$ events are considered to be associated with $K^- \pi^+$ and/or 
$\bar \kappa$ which are in the $\bar{K}^*(892)^0$ region. 
No interference is expected between charge conjugate states.
A coherent $\bar{K}^*(892)^0K^*(1410)$ amplitude is also examined, 
and the difference is also included in the uncertainty for the $\kappa$ parameters. 
Uncertainties for the $\kappa$ parameters contain also
the change from 1$\sigma$ variations of the masses and widths of 
the other resonances and uncertainties of the fit.
Mass and width parameters 
of intermediate resonances 
and of background processes 
in the fit by Method B are also tabulated in Table 2. 
Errors shown in the table are only statistical.

The $\chi^2/$d.o.f value is 1.10.  We examined also the
fit without the $\kappa$ resonance and obtained the value to be
2.83. In the latter, the parameters for 
the $K_S$ background are fixed.

The results obtained in the two analyses reproduce
the data well and are in good agreement with each other.  
Both fits favor strongly that the low mass enhancement of the $K^+\pi^-$
 system  is a resonance. 
The scalar resonance is considered to be the $\kappa$ 
which is necessary in both fits. 
The average values for Breit-Wigner parameters of masses and widths 
for the $\kappa$ (given in the third row in Table 1) 
are obtained from Methods A and B,\\
$$  M_{\kappa} = 878 \pm 23^{+64}_{-55} \ \mbox{MeV}/c^2,\ \  
    \Gamma_{\kappa} = 499 \pm 52^{+55}_{-87} \ \mbox{MeV}/c^2, $$
where the first term errors are statistical ones. The second ones show
total uncertainties, taking the largest values between 
systematic uncertainties of Methods A and B. 
The average values are in good agreement with those obtained  in the analysis
of $K\pi$ scattering phase shifts \cite{sishida1}.
The $\kappa$ parameters obtained are also consistent with those obtained in
the analysis  of $D^+ \rightarrow K^- \pi^+\pi^+$ 
in the E791 experiment \cite{E791} with 
$M_\kappa =797 \pm 19 \pm 43$ MeV/$c^2$ 
and 
$\Gamma_\kappa =410 \pm 43 \pm 87 $ MeV/$c^2$. 
The relative contribution for the 
kappa normalized for the total event number 
ranges between  0.08 and 0.25 taking the 
effects coming from 
the multi-solutions and systematic uncertainties 
in the analyses into account.
%
%
\\

Recent analyses \cite{Mahiko} of $J/\psi \rightarrow 1^-0^-$ decays and
$0^- 0^-$ decays show the large amount of  strong phases 
between relevant amplitudes. 
This fact suggests that, in the relevant $J/\psi \rightarrow  K^*(892)K\pi$ decay, 
the $K\pi$ system is not isolated out of
the final three-body system, and accordingly in the decay amplitude the phase
of the pure $K\pi$-scattering amplitude 
is not directly 
observed experimentally.

In Methods A and B, the phase of the total
$J/\psi \rightarrow  K^*(892)K\pi$ amplitude comes from a sum of respective
contributions of the $S$-matrix elements with the final systems, 
$ K^*(892) \kappa$, $ K^*(892) K_0^*(1430)$, $ K^*(892)K\pi$, etc.
We obtained the $\kappa$ resonance with width, 
$\Gamma_\kappa \simeq 500$MeV/$c^2$ in the Breit-Wigner parameterization, 
which is consistent with the generalized unitarity of S-matrix. 
This behavior is also consistent with the result of analysis of 
$D^+ \rightarrow K^- \pi^+\pi^+$ process in the E791 experiment \cite{E791}.
\\

In conclusion, we have shown that the low mass enhancement in the
invariant  mass spectrum of $K^+ \pi^-$ in the 
$J/\psi \rightarrow \bar K^*(892)^0 K^+ \pi^-$ decays 
comes neither from phase space, nor from other  $J/\psi$ decay processes. 
The angular distribution of $K$ in the $K\pi$ rest frame in the low mass
region shows S-wave decay. 
Two independent analyses for the process, by 
the covariant helicity amplitude method and by 
the VMW method, have been performed, providing a cross check with each other. 
They reproduce the data well, and the results are in good agreement. 
The low mass enhancement is well described 
by the scalar resonance $\kappa$, which is highly required 
in  the analyses.  Parameter values for BW mass and width of the $\kappa$, 
averaged from those obtained by these two methods, are 
878 $\pm$ 23$^{+64}_{-55}$ MeV/$c^2$  and 
499 $\pm$ 52$^{+55}_{-87}$  MeV/$c^2$, 
respectively. 
They are in good agreement with those obtained in the 
analysis on the $K\pi$ scattering phase shifts.
The pole position is determined to be 
$ ( 841 \pm 30^{+81}_{-73} ) - i( 309 \pm 45^{+48}_{-72} )$ MeV/$c^2$
from the average values. \\


{\bf Acknowledgements}\\

The BES collaboration thanks the staff of BEPC for their hard
efforts. This work is supported in part by the National Natural
Science Foundation of China under contracts Nos. 10491300,
10225524, 10225525, 10425523, the Chinese Academy of Sciences under
contract No. KJ 95T-03, the 100 Talents Program of CAS under
Contract Nos. U-11, U-24, U-25, and the Knowledge Innovation
Project of CAS under Contract Nos. U-602, U-34 (IHEP), the
National Natural Science Foundation of China under Contract No.
10225522 (Tsinghua University), the Department of Energy under
Contract No.DE-FG02-04ER41291 (U Hawaii), the Core University
Program of Japan Society for the Promotion of Science, JSPS under
Contract No. JR-02-B4, the fund for the international
collaboration and exchange of RIQS, Nihon-U.


\end{document}